\documentclass[prb, twocolumn, showpacs, amsmath, amssymb,superscriptaddress]{revtex4-1}

\usepackage{amssymb,amsfonts,amsmath} 
\usepackage{graphicx}
\usepackage{color}
\usepackage{hyperref}
\usepackage{longtable}
\usepackage{bbm}
\usepackage{ulem}

\begin{document} 


\title{Transport through Periodically Driven Correlated Quantum Wires}

\author{D.M.\ Kennes}
\affiliation{Department of Physics, Columbia University, New York, NY 10027, USA}

\begin{abstract} 
We study correlated quantum wires subject to harmonic modulation of the onsite-potential concentrating on the limit of large times, where the response of the system has synchronized with the drive. We identify the ratio $\Delta\epsilon/\Omega$ of the driving amplitude $\Delta\epsilon$ and the frequency of driving $\Omega$ as the scale determining the crossover from a modified Luttinger liquid picture to a system that behaves effectively like a higher dimensional one. We exemplify this crossover by studying the frequency dependency of the boundary density of state $\rho_{\rm B}(\omega)$ as well as the temperature dependency of the linear conductance $G(T)$ through the wire, if the latter is contacted to leads. Both observables are known to exhibit Luttinger liquid physics without driving given by characteristic power-law suppression as $\omega\to\epsilon_{\rm F}$ (with $\epsilon_{\rm F}$ the Fermi energy) or $T\to 0$, respectively. With driving we find that this suppression is modified from a single power-law to a superposition of an infinite number of power laws. At small $\Delta\epsilon/\Omega\ll 1$ only a few terms of this infinite sum are relevant as the prefactors of higher terms are suppressed exponentially. Thus a picture  similar to the equilibrium Luttinger liquid one emerges. Increasing  $\Delta\epsilon/\Omega$ an increasing number of power laws contribute to the sum and approaching $\Delta\epsilon/\Omega\gg 1$ the system behaves effectively two dimensional, for which the suppression is wiped out completely.

\end{abstract}

\date{\today} 
\maketitle

\section{Introduction}
With the advent of ultra-cold gases\cite{Bloch08} as well as due to significant improvements in the techniques used to manipulate solids,\cite{Yu91,Miyano97,Wall11,Blumenstein11} controllable non-equilibrium experiments of condensed matter systems are accessible nowadays. The case of periodic external driving is  experimentally routinely realized\cite{Lignier07,Kierig08,Sias08} rendering a better understanding of the underlying non-equilibrium physics in such  systems imperative.  

In the context of periodically driven systems the so called Floquet matter has recently attracted an immense amount of research activity.\cite{Else16,Khemani16a,Keyserlingk16,Khemani16b,Moessner17}  Intriguingly, in the periodically driven case  novel non-equilibrium phases are realized. Analogous to equilibrium phases, those can be classified by local order parameters for which the discrete time-translation symmetry which is generated by the external drive is spontaneously broken. Certain periodically driven system thus provide examples of  so called 'time crystals', which were first hypothesized for undriven systems in Ref. \onlinecite{Wilczek}. However, in closed driven interacting systems these novel phases of matter might be difficult to realize as generic systems risk to heat up to an infinite temperature state (at which all interesting phases are lost), because the drive continuously increases the energy in the system in the presence of interactions. As a mechanism to avoid this heating many-body localization has been suggested and first promising experiments exploiting this route have been performed recently.\cite{Bordia17,Zhang17,Choi17} Another route to avoid heating is to add reservoirs which can absorb the additional energy put into the system by the drive. 
In the context of Floquet matter this has been suggested,\cite{Else17} but not extensively studied so far. Besides of time crystals, periodic driving was also identified to allow to control intriguing topological properties.\cite{FTI} Corresponding systems were dubbed Floquet topological insulators.  Yet another interesting application of external periodic drive is the control of quantum dots.\cite{QDp,Eissing16a,Eissing16b}  
 
Here, we investigate a one-dimensional quantum wire of interacting fermions subject to a periodic drive.\cite{LLperio} We aim to contribute to the understanding of periodically driven open systems and thus we consider in addition  weakly coupled reservoirs of particles, which can act as  heat baths absorbing access energy put in by the drive. We exclusively consider the limit of large times (the drive has been turned on in the far past) where the system has synchronized with the external drive. 

One-dimensional fermionic systems, such as the ones in the focus of this work, exhibit an interesting signature of strongly correlated physics: the breakdown of perturbative approaches, which treat all energy scales at once. This breakdown is signaled by logarithmic divergences in physical quantities such as the mode occupancy $n_k$, the local spectral function $\rho(\omega)$ or the conductance $G(T)$ through the wire, if the latter are end-contacted by leads.  To extract physically meaningful results renormalization group approaches can be employed, in which energy scales are treated successively from high to low. Employing these RG approaches one succeeds in resuming the logarithmic divergences found in perturbative approaches to their correct power-law form. One of these renormalization group methods is the so-called functional renormalization group (FRG).\cite{Kopietz10,Metzner12} This method has the advantage that microscopic models can be treated directly (instead of field theories) gaining access to the physics on all energy scales. In applications the FRG requires the truncation of the set of RG flow equations which can be justified for weak to intermediate two-particle interactions. In the context of one-dimensional quantum wires it has been shown before\cite{Meden02,Andergassen04,Enss05} that lowest order truncated FRG leads to power-law behavior of boundary and impurity physics, with the exponent being correct to leading order in the interaction. 
Later FRG\cite{Jakobs07,Kennes13} was applied to quantum wires out of equilibrium and an interesting modification of the power-laws by the applied bias voltage has been identified (see below). The interplay of strong correlations as manifest, e.g., in the suppression of the spectral function at a boundary of a one dimensional wire or the conductance, and periodic driving is thus an intriguing generalization, for which we will employ the methodological advances put forward in Refs.~\onlinecite{Eissing16a,Eissing16b} combined with the, for extended systems, more suitable RG cutoff introduced in Ref.~\onlinecite{Jakobs07}.  

Compared to equilibrium, the  periodic driving introduces an additional control knob for tuning the Luttinger liquid physics. In  equilibrium the power-law suppression of the boundary spectral function $\rho_{\rm B}(\omega)$ follows $|\omega-\epsilon_{\rm F}|^{\alpha_{\rm B}}$  at temperature $T=0$  (with known\cite{Meden02,Andergassen04,Enss05} equilibrium boundary exponent $\alpha_{\rm B}$). Our main result is that with driving the time averaged boundary spectral function shows suppressions in frequency space separated by integer multiples of the driving frequency $\Omega$. For each of these, the suppressions of the spectral weight, time-averaged over one period, follows not a single (like in equilibrium) but a superposition of power-laws
\begin{align}
&\rho_{\rm B}(|\omega-\epsilon_{\rm F}-r\Omega|)\stackrel{|\omega-\epsilon_{\rm F}-r\Omega|\ll 1}{\approx}\notag\\& \sum_{r'=-\infty}^{\infty} c_{r,r'} \left|J_{r'}\left(\frac{2\Delta\epsilon}{\Omega}\right)\right|^2  |\omega-\epsilon_{\rm F}-r\Omega|^{\left|J_{r-r'}\left(\frac{2\Delta\epsilon}{\Omega}\right)\right|^2\alpha_{\rm B}},\label{eq:sup_powerlaw_intro}
\end{align}
where the coefficients $c_{k,k'}$ depend on the details of the underlying microscopic model, $r$ is an integer and $J_r(x)$ is the $r$-th Bessel function of the first kind. 

Importantly, this behavior of the time-averaged spectral function, manifests in transport properties such as the time averaged linear conductance $G(T)$.  If we consider a single impurity in the center of the wire end-contacted addiabatically to non-interacting reservoirs (see below for a precise definition)
we find \begin{equation}
G(T)\sim \sum\limits_{r=-\infty}^\infty \left|J_{r}\left(\frac{2\Delta\epsilon}{\Omega}\right)\right|^2 \left(\frac{T}{T_0}\right)^{2 \left|J_{r}\left(\frac{2\Delta\epsilon}{\Omega}\right)\right|^2 \alpha_B}
\end{equation}
while we obtain
\begin{equation}
G(T)\sim \sum\limits_{r=-\infty}^\infty \left|J_{r}\left(\frac{2\Delta\epsilon}{\Omega}\right)\right|^2 \left(\frac{T}{T_0}\right)^{ \left|J_{r}\left(\frac{2\Delta\epsilon}{\Omega}\right)\right|^2 \alpha_B}
\end{equation}
in the case of an abruptly end-contacted quantum wire without an additional impurity in its center. The single power-law suppression of the equilibrium, undriven case (given by the above equations in the limit $\Delta \epsilon \to 0$, such that $J_r(2\Delta\epsilon/\Omega)=\delta_{r,0}$) is replaced by an infinite sum of power-laws. At small to intermediate $\Delta \epsilon/\Omega$ only a few of these power-laws contribute appreciably and a modified Luttinger liquid picture emerges. In the opposite limit of $\Delta \epsilon/\Omega\gg 1$ all addents of the infinite sum of power-laws are relevant and sum up to a regular contribution which wipes out the suppression completely.

The rest of this paper is organized as follows. In the second section we outline the time dependent model being studied. Section \ref{sec:FRG} depicts the functional renormalization group treatment used to obtain the results presented in Sec.~\ref{sec:results}. The last section gives a concluding summary.

\begin{figure}[t]
\centering
\includegraphics[width=\columnwidth]{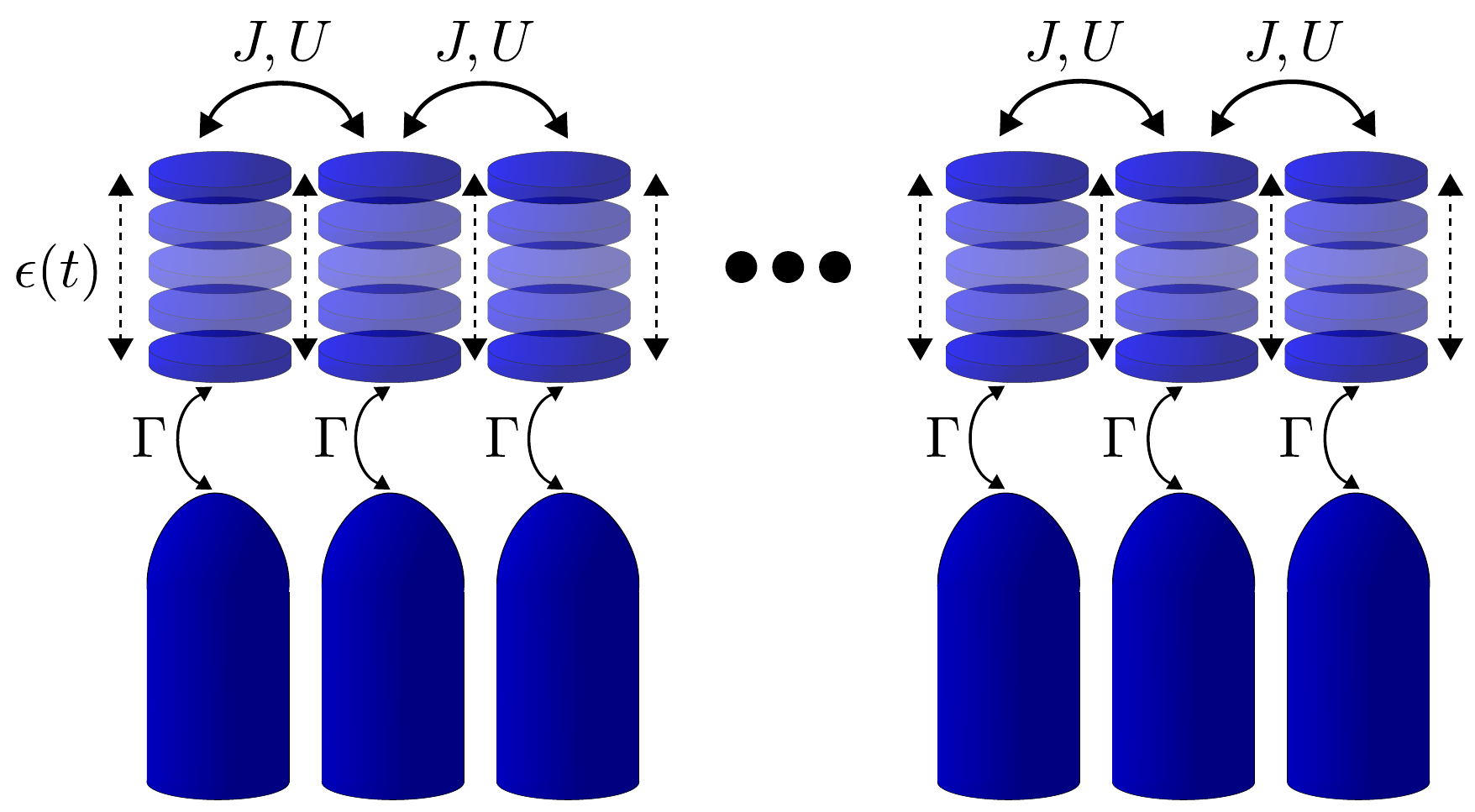}
\caption{The studied model consists of a chain of lattice sites harboring spinless fermions. The chain is characterized by a nearest neighbor hopping $J$ as well as a nearest neighbor density-density interaction $U$. Additionally the onsite energies are subject to periodic change $\epsilon(t)=2\Delta\epsilon\cos(\Omega t)$. We concentrate on this simple form of driving to gain analytic insights into the problem, but more complicated forms of drivings can be tackled numerically within the presented framework. The influence of a thermodynamic reservoir is modeled by individual structureless reservoirs inducing the hybridization $\Gamma$. In the main text we focus on two cases: (i) every site is coupled to an individual reservoir (in Sec.~\ref{subsec:Spec}) as well as (ii) only the first and last sites are coupled to a reservoir (in Sec.~\ref{subsec:Cond}). } 
\label{fig:Model}
\end{figure}

\section{Model}
\label{sec:model}
The system under scrutiny describes spinless fermions in a one-dimensional lattice, with nearest neighbor hopping and interaction, 
\begin{align}
H^{\rm S}=&\sum_{i=1}^{N-1} -J_i \left(c^\dagger_ic_{i+1} +c^\dagger_{i+1}c_i \right)\notag\\
&+U_i \left(n_i-\frac 12\right) \left(n_{i+1}-\frac 12\right)+2\Delta\epsilon \cos(\Omega t)\sum_{i=1}^{N}n_i.
\end{align} 
The operators $c^{(\dagger)}_i$ is the annihilation (creation) operator of a fermion on site $i$ and the density $n_i=c^\dagger_i c_i$. Each site can be coupled to its own reservoir, such that the total Hamiltonian reads
\begin{equation}
H=H^{\rm S}+H^{\rm T}+H^{\rm Res},
\end{equation} 
where the terms
\begin{equation}
H^{\rm T}=\sum_{i=1}^N\int dk_i t_{i,k_i}c^\dagger_i a_{i,k_i}+{\rm H.c.}
\end{equation}  and 
\begin{equation}
H^{\rm Res}=\sum_{i=1}^N\int dk_i \epsilon_{k_i} a_{i,k_i}^\dagger a_{i,k_i}
\end{equation}
describe this system-reservoir coupling as well as the reservoir, respectively. For our purposes we want to neglect the structural details of the reservoirs and use a wide-band description: $\epsilon_{k_i}=v_i k$ and $t_{i,k_i}=t_i$. For wide bands the reservoirs introduce an energy independent, but site dependent hybridization of strength $\Gamma_i=2\pi |t_{i}|^2/v_i $. The  cases of interest to this study are: (i) site independent hybridisation $\Gamma_i=\Gamma$ as well as (ii) an end-contacted wire $\Gamma_1=\Gamma_N=\Gamma$ and $\Gamma_{1<i<N}=0$. For simplicity we will focus on reservoirs with zero chemical potential $\mu=0$ and common temperature $T$. Additionally, we concentrate on the, to us, most interesting case of the coupling $\Gamma_i$ being sufficiently small such that the presence of the reservoirs does not completely destroy the Luttinger liquid physics of the isolated system $H^{\rm S}$. In the opposite limit $\Gamma$ would provide a large cutoff for the power-laws (which we take as the hallmark of a Luttinger liquid) found in observables and thus wipe out the Luttinger liquid physics of interest to us. The system is depicted in Fig.~\ref{fig:Model}.

We allow for site dependent hopping amplitudes $J_i$ to study the effects arising from the introduction of impurities on the transport properties of the driven wire. In these transport set-ups we will concentrate on the case (ii) of an end-contacted wire for the reservoirs described in the previous paragraph. We introduce the notion of a bulk hopping amplitude $J$, which sets the bandwidth of the underlying homogeneous wire, such that away from the impurity $J_i=J$. Site dependent interactions $U_i$ will be used to adiabatically connect to the non-interacting wires to minimize the scattering effects induced by the contacts, where needed (see below). 

To simplify the following discussion we will concentrate on the case, where the bandwidth (related to $J$) of the wire is large compared to the other energy scales, such that the influence of the curvature of the band can be disregarded. In the opposite limit one would have to include the effects of the band curvature onto the boundary exponents, which though straightforward in our approach, constitutes an unnecessary complication to the physics we want to address.

\section{FRG Treatment} 
\label{sec:FRG}
We use a combination of the techniques outlined in detail in Refs.~\onlinecite{Eissing16a}, \onlinecite{Eissing16b} and \onlinecite{Jakobs07} to treat the asymptotic long time behavior of the system described above. We thus access directly the state at large times, where the response of the quantum chain has synchronized with the drive.

Treating the interacting many-body quantum system described in Sec.~\ref{sec:model} within the FRG can be broken down into to a three-step procedure using the language of Green's functions:
\begin{itemize}
\item calculate the Floquet Green's functions of the non-interacting, decoupled wire.
\item determine the non-interacting, reservoir-dressed  Floquet Green's functions, by including the self-energy induced by the reservoirs.
\item utilize the FRG to describe the effects of interactions, which yields an approximation to the interacting Green's function.  
\end{itemize}
Each of the following subsections is devoted to one of these separate steps.

\subsection{Non-interacting, decoupled Green's functions}
The non-interacting, decoupled Green's function can be calculated from the single particle Floquet Hamiltonian,\cite{Eissing16a,Eissing16b} which reads
\begin{align}
H^{\rm S, eff}=\begin{pmatrix}
\ddots&D(\Delta\epsilon)&0&0&0\\[6pt]
D(\Delta\epsilon)&A(1)&D(\Delta\epsilon)&0&0\\[6pt]
0&D(\Delta\epsilon)&A(0)&D(\Delta\epsilon)&0\\[6pt]
0&0&D(\Delta\epsilon)&A(-1)&D(\Delta\epsilon)\\[0pt]
0&0&0&D(\Delta\epsilon)&\ddots\\[6pt]
\end{pmatrix}
\label{eq:Heff_outer}
\end{align}
with 
\begin{equation}
A(n)=\begin{pmatrix}
n\Omega & -J_1 & & 0\\
-J_1 & \ddots & \ddots &  \\
& \ddots & \ddots &  -J_{N-1} \\
0 &  & -J_{N-1} & n\Omega  \end{pmatrix}
\label{eq:Heff_inner}
\end{equation} 
and $D(\Delta\epsilon)=\Delta\epsilon\mathbbm{1}$, where $\mathbbm{1}$ is the $N$ dimensional unity matrix.
The Hamiltonian is thus a matrix in Floquet space with $n,n'$ [the indices of the matrix in Eq.~\eqref{eq:Heff_outer}] as well as (site) indices $i,i'$ [the indices of the matrices in Eq.~\eqref{eq:Heff_inner} and $D(\Delta\epsilon)$]. Since $D(\Delta\epsilon)\sim \mathbbm 1$ the effective Hamiltonian is easily diagonalized by first diagonalizing  with respect to the site indices $i,i'$ 
\begin{widetext}
\begin{align}
\begin{pmatrix}\ddots&&&&\\
&U^{-1}&&&\\
&&U^{-1}&&\\
&&&U^{-1}&\\
&&&&\ddots
\end{pmatrix}
H^{\rm S, eff}
\begin{pmatrix}\ddots&&&&\\
&U&&&\\
&&U&&\\
&&&U&\\
&&&&\ddots
\end{pmatrix}
=\begin{pmatrix}
\ddots&D(\Delta\epsilon)&0&0&0\\[6pt]
D(\Delta\epsilon)&d(1)&D(\Delta\epsilon)&0&0\\[6pt]
0&D(\Delta\epsilon)&d(0)&D(\Delta\epsilon)&0\\[6pt]
0&0&D(\Delta\epsilon)&d(-1)&D(\Delta\epsilon)\\[0pt]
0&0&0&D(\Delta\epsilon)&\ddots\\[6pt]
\end{pmatrix}
\end{align}
\end{widetext}
with $d(l)_{k,k'}=\delta_{k,k'}[l\Omega+\epsilon(k)]$
and then diagonalizing the $N$ independent infinite Wannier-Stark ladders\cite{Hacker70} shifted only by the global energy $\epsilon(k)$.
The unitary transformation $R$ which 
generates the second diagonalization step is known analytically  $R_{n,l}=(-1)^{n-l}J_{n-l}\left(\frac{2\Delta\epsilon}{\Omega}\right)$.

The next subsection includes explicit equations for the reservoir dressed Green's functions building on the diagonalized Hamiltonian. The case of the non-interacting, reservoir-decoupled Green's functions is included as a special case ($\Gamma_i\to 0$) and we refrain from giving the (less general) equations here.

\subsection{Non-interacting, reservoir-dressed Green's functions}
For the reservoir dressed retarded and advanced Green's functions the above diagonalization scheme needs to be modified slightly, because the reservoir retarded and advanced self-energy are, in general, site-dependent. The retraded and advanced self-energy induced by the reservoirs reads
 \begin{equation}
 \Sigma^{\rm Ret/Adv}_{n,i,n',i}=\delta_{n,n'}\Sigma^{\rm Ret/Adv}_{i,i}=\mp i\frac{\Gamma_i}{2} \delta_{n,n'}\delta_{i,i'}.
 \end{equation} 
 Defining $\tilde h(n)=A(n)+\Sigma^{\rm Ret/Adv}$ and assuming that a set of left and right eigenvectors $\{\langle \hat k|\}$ and $\{\left|k\right\rangle\}$ of $\tilde h(0)$ exist,\cite{note1} we find that these are also left and right eigenvectors of $\tilde h(n)$, with
\begin{align*}
\tilde h(n)\left|k\right\rangle&=(\lambda_k+n\Omega)\left|k\right\rangle\;\;\;\;\;\;\;\;\tilde h(n)^\dagger|\hat k\rangle&=(\lambda_k^*+n\Omega)|\hat k\rangle\\
\langle \hat k|\tilde h(n)&=(\lambda_k+n\Omega)\langle\hat k|\;\;\;\;\;\;\;\;\langle k|\tilde h(n)^\dagger&=(\lambda_k^*+n\Omega)\langle k|
\end{align*}
and
\begin{equation}
\sum_k|\hat k\rangle\langle k|=\sum_k|  k\rangle\langle \hat k|=1.
\end{equation}
Again, diagonalizing with respect to the Floquet index using $R$ (see above) the Green's functions can be expressed using the eigenstates $|l\rangle$ of the Wannier-Stark ladder 
\begin{align}
G^{\rm Ret}(\omega)=\sum_{l,k}\frac{1}{\omega-(\lambda_k+l\Omega) }|  k,l\rangle\langle \hat k,l|,\\
G^{\rm Adv}(\omega)=\frac{1}{\omega-(\lambda_k^*+l\Omega)  }|\hat k,l\rangle\langle k,l|.
\end{align}
We can easily transform back into the original basis $(n,i)$ by inverting the above transformations.

The lesser (or Keldysh) component of the Green's function is slightly more complicated to compute, because the reservoir lesser self-energy is given by $\Sigma^{<}_{n,i,n',i}(\omega)= \Gamma_i f(\omega-n\Omega) \delta_{n,n'}\delta_{i,i'}$ with  $f(\omega)=1/(e^{\omega/T}+1)$ the Fermi-Dirac distribution and the self-energy is thus not $\sim \mathbbm{1}$ in the Floquet space. We define $f^r(\omega)=f(\omega-r\Omega)$ and use  
\begin{align}
G^{<}(\omega)=&-G^{\rm Ret}\Sigma^<G^{\rm Adv}\notag\\
=&\sum_{l,k,l',k'}\frac{1}{\omega-(\lambda_k+n\Omega) }|  k,l\rangle\langle \hat k,l|\Sigma^< |\hat k',l'\rangle\langle k',l'|\notag\\&\phantom{\sum_{l,k,l',k'}}\frac{1}{\omega-(\lambda_{k'}^*+l'\Omega)  }\notag\\
=& -\left(G^{\rm Ret} N-N G^{\rm Adv}\right),
\end{align}
where 
\begin{equation}
N(\omega)=\sum_{l,k,l',k'}|  k,l\rangle \frac{\langle \hat k,l|\Sigma^< |\hat k',l'\rangle}{\lambda_k-\lambda_{k'}^*+(l-l')\Omega}\langle k',l'|
\end{equation}
is an effective distribution operator. The lesser self-energy can be written as
\begin{widetext}
\begin{align}
\langle\hat k,l|\Sigma^<|\hat k',l'\rangle=\sum_{r=-\infty}^\infty f^r(\omega) (-)^{l-l'}J_{r-l}\left(\frac{2\Delta\epsilon}{\Omega}\right)  J_{r-l'}\left(\frac{2\Delta\epsilon}{\Omega}\right)\langle \hat k|\lambda_k-\lambda_{k'}^* |\hat k'\rangle
\end{align}
yielding 
\begin{equation}
N(\omega)=\sum_{r=-\infty}^\infty f^r(\omega) \sum_{l,k,l',k'}(-)^{l-l'}J_{r-l}\left(\frac{2\Delta\epsilon}{\Omega}\right)  J_{r-l'}\left(\frac{2\Delta\epsilon}{\Omega}\right) |  k,l\rangle \frac{\langle \hat k|\hat k'\rangle}{1+(l-l')\frac{\Omega}{\lambda_k-\lambda_{k'}^*}}\langle k',l'|\label{eq:Nbfdia}
\end{equation}
\end{widetext}
Because $\{\langle \hat k|\}$ and $\{\left|k\right\rangle\}$  are left and right eigenvectors of a non-hermitian matrix $\tilde h(n)=A(n)+\Sigma^{\rm Ret/Adv}$, they are  not (necessarily) orthogonal and $\langle \hat k|\hat k'\rangle$ is in general not $\sim\delta_{k,k'}$. However, the non-Hermitian nature of this matrix is rooted only in $\Sigma^{\rm Ret/Adv}$ with contributions $\sim\Gamma_i$ adding an imaginary part to the otherwise Hermitian matrix $A(n)$. Since we are only interested in small system-reservoir couplings  ${\rm max}(\Gamma_i)\ll \Omega$, one can show that to leading order in $\Gamma_i/\Omega$, $\langle \hat k|\hat k'\rangle=\delta_{k,k'}+\mathcal{O}(\Gamma_i/\Omega)$. Of course in the special case $\Gamma_i=\Gamma$, $\langle \hat k|\hat k'\rangle= \delta_{k,k'}$ exactly, because then $|\hat k\rangle=|k\rangle$.

Using $\langle \hat k|\hat k'\rangle=\delta_{k,k'}+\mathcal{O}(\Gamma_i/\Omega)$ we can re-express\cite{Jakobs07} Eq.~\eqref{eq:Nbfdia} to lowest in $\Gamma_i/\Omega$
\begin{equation}
N(\omega)=\sum_{r=-\infty}^\infty f^r(\omega)\sum_{l,k}\left| k,l\right\rangle \left|J_{r-l}\left(\frac{2\Delta\epsilon}{\Omega}\right)\right|^2 \langle \hat k,l|.\label{eq:eff_dis}
\end{equation}    
In the limit $\Gamma\ll \Omega$ the effective distribution operator is approximately diagonal in the same basis as the retarded and advanced Green's functions and as a consequence so is $G^<$. 

We want to interpret this result and its consequences in a little more detail. We only assume that the hybridization is small, but treat generic ratios of $\Omega$ and $\Delta \epsilon$. We are therefore {\it not} restricted to the adiabatic limit ${\rm min}(\Gamma_i)\gg \Omega$, where the system of interest would have enough time to follow the external parameter change at each instance in time. 
Because we do treat arbitrary ratios of $\Omega$ and $\Delta \epsilon$, our study includes, but goes beyond, the high-frequency limit $\Omega \gg \Delta \epsilon$, where the influence of the drive  would vanish (as it averages to zero) in accordance to a lowest order Magnus expansion.\cite{Magnus64} Thus we can access the interesting regime beyond the adiabatic and high frequency limit and how  the physics of such a system subject to driving with general frequency eventually crosses over to its high-frequency behavior upon increasing the driving frequency. Furthermore, we highlight an intriguing consequence of the simple form of the effective distribution function Eq.~\eqref{eq:eff_dis} in the considered limit ${\rm max}(\Gamma_i)\ll \Omega$: frequency integrated single particle observables take a particularly simple form which is given below. Frequency integrated quantities include important examples, such as the occupancies and the particle currents. Moreover, they also determine the effective potentials generated in first order truncated FRG employed in this paper (see below).\cite{note2} 

To understand this we explicitly consider the frequency integrated lesser Green's functions, which reads
\begin{widetext}
\begin{align}
\frac{1}{2\pi}\int d\omega G^{<}(\omega)=&-\frac{1}{2\pi}\int d\omega\sum_{r=-\infty}^\infty\sum_{l,k}\left(\left| k,l\right\rangle  f^r(\omega)\left|J_{r-l}\left(\frac{2\Delta\epsilon}{\Omega}\right)\right|^2 \frac{1}{\omega-(\lambda_k+l\Omega) }\langle \hat k,l|\right.-{\rm H.c.}\bigg)\notag\\
=&-\frac{1}{2\pi}\sum_{l,k}\sum_{r=-\infty}^\infty\int d\omega\left(\left| l,k\right\rangle  f^{r+l}(\omega)\left|J_{r}\left(\frac{2\Delta\epsilon}{\Omega}\right)\right|^2  \frac{1}{\omega-(\lambda_k+l\Omega) }\langle \hat k,l|-{\rm H.c.}\right)\notag\\
=&-\frac{1}{2\pi}\sum_{l,k}\sum_{r=-\infty}^\infty\int d\omega\left(\left| l,k\right\rangle  f^{r}(\omega)\left|J_{r}\left(\frac{2\Delta\epsilon}{\Omega}\right)\right|^2  \frac{1}{\omega-\lambda_k }\langle \hat k,l|-{\rm H.c.}\right)\notag\\
=&-\frac{1}{2\pi}\sum_{r=-\infty}^\infty\int d\omega\left(  f^{r}(\omega)\left|J_{r}\left(\frac{2\Delta\epsilon}{\Omega}\right)\right|^2  \sum_{l,k}\left| l,k\right\rangle\frac{1}{\omega-\lambda_k }\langle \hat k,l|-{\rm H.c.}\right)\notag\\,
\end{align}  
\end{widetext}
where we performed a shift $r\to r+l$ in the second  and    $\omega\to \omega + l\Omega$ in the third line. Importantly this shows that the lesser Green's function is not only diagonal, but also proportional to unity in the Floquet index 
$l$. Rotating back to the original space of space $i$ and $n$ this entails
 \begin{widetext}\begin{equation}
\frac{1}{2\pi}\int d\omega G^{<}(\omega)_{n,i,n',i'}=-\delta_{n,n'}\frac{1}{2\pi}\sum_{r=-\infty}^\infty\int d\omega  f^{r}(\omega)\left|J_{r}\left(\frac{2\Delta\epsilon}{\Omega}\right)\right|^2 \left(\tilde G^{\rm Ret}_{i,j}(\omega)  -{\rm H.c.}\right),
\end{equation}\end{widetext}
where $\tilde G^{\rm Ret}_{i,j}$ is the retarded Green's function calculated with respect to the undriven single-particle Hamiltonian characterized by the dispersion $\epsilon(k)$ and the retarded reservoir self-energy.  $\tilde G^{\rm Ret}_{i,j}$ can be interpreted physically as the Green's function in the rest frame of the wire.  Importantly, thus the limit $\Gamma_i\ll\Omega$ leads to a lesser Green's function which is diagonal in the Floquet space $n,n'$. In time-space this means that only the time averaged component has finite expectation value. The above result shows that the calculation of the frequency integrated Floquet lesser Green's function can be simplified to a similar calculation without the Flouqet index (or to be more precise $n$ of these calculations with shifted chemical potential $\mu\to \mu+n\Omega$), which reduces the matrix dimension (and therefore complexity of the problem) immensely.

\subsection{Interacting Green's functions: FRG}

Finally, we include the interactions via the FRG approach. We use the simplest/lowest order truncation scheme of the FRG hierarchy of flow equations in which the self-energy acquires a flow, but the effective two-particle interaction vertex is set to its bare value. This approximate truncation scheme was shown to describe crucial aspects of the boundary and impurity physics of quantum wires accurately in-\cite{Meden02,Andergassen04,Enss05} as well as out-of-equilibrium.\cite{Jakobs07} However, inelastic scattering is neglected and the intriguing question of the competition of heat production by these processes and taking out heat by the external reservoirs can consequently not be addressed in a meaningful manner employing this approximation (the effects are of higher order in the interactions). 


We employ the cut-off scheme proposed in Ref.~\onlinecite{Jakobs07} in which the flow parameter $\Lambda$ is introduced within the distribution function of the reservoirs
\begin{equation}
f^{r}(\omega)\to f^{r,\Lambda}(\omega)=T \sum_{\omega_n}\frac{\Theta(|\omega_n|-\Lambda)}{i\omega_n-\omega+r\Omega}, 
\end{equation}
with  $\omega_n=(2n+1)\pi T$ being the Matsubara frequencies and $\Theta(\omega)$ the Heaviside-Theta function. Within the lowest order truncated FRG approach the flow equation of the self-energy describes the renormalization of the Hamiltonian parameters. Therefore the results derived above can be directly applied to the interacting case if (a) the renormalized $\Delta\epsilon$ remain site independent and (b) the hybridization remains small. The renormalization of the different components of the effective single-particle Floquet Hamiltonian are proportional to the $\omega$-integrated single-scale propagator at Floquet indeces $(n,0)$\cite{Eissing16a,Eissing16b}
\begin{align}
S^{{\rm Ret/Adv},\Lambda}(\omega)&=0\\
\int d\omega \left.S^{<,\Lambda}(\omega)\right|_{n,i,0,i'}&=\delta_{n,0}\sum_{r}
\left|J_{r}\left(\frac{2\Delta\epsilon}{\Omega}\right)\right|^2 \notag\\
\times\tilde G^{\rm Ret}_{i,i'}&(i\omega_{n(\Lambda)}-r\Omega)+ ({{\rm Ret}\leftrightarrow {\rm Adv}}),
\end{align}  
where $\omega_{n(\Lambda)}$ is the Matsubara frequency closest to $\Lambda$. Thus only the $n=0$ (time-averaged) component of the Floquet Hamiltonian is renormalized by the interaction. The total flow reads
\begin{align}
\frac{\partial}{\partial \Lambda}&\left.\Sigma_U^{{\rm Ret/Adv},\Lambda}\right|_{n,i,0,i'}=\notag\\&-\delta_{n,0}T\sum_{r,\omega_n,i_1,i_2}\delta(|\omega_n|-\Lambda)\bar{v}_{ii_1,i'i_2}
\notag\\&\;\;\;\;\;\;\;\;\;\;\;\;\;\times\left|J_{r}\left(\frac{2\Delta\epsilon}{\Omega}\right)\right|^2 
 \tilde G^{\rm Ret}_{i_2,i_1}(i\omega_{n(\Lambda)}-r\Omega)\notag\\&+ ({{\rm Ret}\leftrightarrow {\rm Adv}})\label{eq:SigmaU1}\\
\frac{\partial}{\partial \Lambda}&\Sigma_U^{ <,\Lambda}=0\label{eq:SigmaU2}
\end{align} 
with $\bar{v}_{ii_1,i'i_2}$ the anti-symmetrized two-particle interaction. 
Therefore, the flow renormalizes only the mean (time-averaged) component and thus (a) holds. Because the interaction is local to the wire the hybridization does not flow at all and thus (b) holds as well. With this the above procedure of diagonalizing the effective Floquet Hamiltonian remains valid even throughout the flow.

We note that in the presence of periodic driving and within the lowest order truncation scheme used here, the flow of the retarded non-equilibrium self-energy $\Sigma_U^{{\rm Ret/Adv}}$ is given by a weighted sum $\sum_r$ of equilibrium flows (compare, e.g., Ref.~\onlinecite{Enss05}), with weighting factor $\left|J_{r}\left(\frac{2\Delta\epsilon}{\Omega}\right)\right|^2$ and shifted chemical potential $\mu\to\mu-r\Omega$. For reference the equilibrium flow can be obtained in the limit $\Delta\epsilon\to 0$. This is similar to the result found in the same truncation in the bias voltage driven case.\cite{Jakobs07} Exploiting this connection  will help us to analyze the structure of the flow equation in more depth and to even derive analytic predictions for the periodically driven case. To keep this paper self-contained we will recapitulate the main findings in equilibrium and how they relate to the driven case in the following subsection. 

All plots are obtained by solving Eq.~\eqref{eq:SigmaU1} numerically without any further approximations. We compare these numerical results, whenever possible, to our analytic predictions (derived in Sec.~\ref{sec:results}) which involve additional simplification valid in $\mathcal{O}(U^2)$.

\subsection{equilibrium results and relation to the periodically driven system}

The flow Eqs.~\eqref{eq:SigmaU1} and \eqref{eq:SigmaU2} are given by a weighted sum $\sum_r$ of equilibrium flows. This allows to adapt parts of the intuition established from the well-understood equilibrium case to the case of periodic driving studied in this paper. 

In equilibrium ($\Delta \epsilon\to 0$ limit in the model defined above) the changes in the effective hopping amplitude $\left.\Sigma^{\rm Ret}_U\right|_{i,i+1}$ (there is no Floquet index in equilibrium) due to the renormalization, are well documented.\cite{Meden02,Andergassen04,Enss05,Metzner12} These are the only non-zero matrix elements of $\Sigma^{\rm Ret}_U$ for the particle-hole symmetric Hamiltonian considered here. Interactions in the wire lead to a band broadening (average value of the hopping increases). If additionally a single impurity (either a link of reduced strength in the wire or the end-contact between the interacting and the non-interacting part of the system) is introduced the effective hopping $J_i+\left.\Sigma^{\rm Ret}_U\right|_{i,i+1}$ exhibit oscillations with wave number $2k_{\rm F}$
and a decaying amplitude. Sufficiently away from the impurity the amplitude of the oscillations decay as $1/|j-j_0|$  at $T=0$, with $j_0$ the position of the impurity.\cite{Meden02} At finite $T$ this power-law scaling holds only up to a thermal length scale $\sim 1/T$ beyond which the decay turns exponential.\cite{Andergassen04,Enss05} Calculating  the scattering off such a long-ranged oscillatory potential (so-called Wigner-von Neumann potential),\cite{Reed75}  leads to the power-law suppression of the conductance and the local spectral function at low temperatures or frequencies, respectively. One can show analytically that the prefactor of
the asymptotic $1/|j-j_0|$ decay determines the exponent $\alpha_{\rm B}$ of this power-law suppression.\cite{Barnab05}  The first order truncated FRG
flow equations yield an amplitude of this decay, and in turn an exponent, which is  independent of the impurity strength,\cite{Meden02} which is the expected behavior.\cite{Kane92} The magnitude of the exponent controlling this suppression is known to agree to the field theoretical results to leading order in $U$ within the lowest order FRG scheme.  The equilibrium exponent $\alpha_{\rm B}=1/K-1$ of the suppression of the local density of states can be related to the so-called Luttinger parameter $K$. Using Bethe ansatz one can calculate $K$ exactly for our model of spinless fermions
\begin{equation}
K=\frac{\pi}{2\arccos\left(-\frac{U}{2J}\right)}\label{eq:K}.
\end{equation}
The scaling of the conductance is more involved.\cite{Kane92} For a single strong impurity far away from the boundary of the wire the conductance vanishes as a power-law in the temperature with exponent $2\alpha_{\rm B}$. This scaling changes to $\alpha_{\rm B}$ when the strong impurity is placed at the boundary (e.g. by abruptly coupling the interacting wire to non-interacting leads). This can be explained within a simple tunneling picture. If we consider two weakly coupled interacting wires, particles have to tunnel from one wire showing a power-law suppression in the boundary spectral function to another one with the same suppression (hence the scaling with two times the exponent $\alpha_{\rm B}$). However, if we consider two weakly coupled wires of which only one is interacting and the other one in non-interacting, particles tunnel from one wire showing a power-law suppression in the boundary spectral function to one wire, which does not exhibit such a power-law suppression (hence the scaling with one times the exponent $\alpha_{\rm B}$).

Returning to the periodically driven case we repeat that each of the Floquet channels $r$ (in the sum $\sum_r$) in the flow equations is weighted by $\left|J_{r}\left(\frac{2\Delta\epsilon}{\Omega}\right)\right|^2$ and shifted in chemical potential by $r\Omega$ w.r.t. to the equilibrium flow. Therefore, each of these channels generates oscillations in the effective hopping, with wave number $2k_{{\rm F},r}$, where $k_{{\rm F},r}$ is the Fermi wave number with respect to the shifted chemical potential (depending on $r$) and an amplitude (and in turn boundary exponent) weighted by $\left|J_{r}\left(\frac{2\Delta\epsilon}{\Omega}\right)\right|^2$. Due to the feedback of the self-energy to its own flow-equation these oscillations generated by one of the Floquet channels will influence the formation of the oscillations of all the other Floquet channels. However, this mutual influence is of order $\mathcal{O}(U^2)$ and will be disregarded to make analytic progress. This additional simplification of the flow equations of order $\mathcal{O}(U^2)$ is carefully checked to hold by comparing to the full numerics in the following. 

\begin{figure}[t]
\centering
\includegraphics[width=\columnwidth]{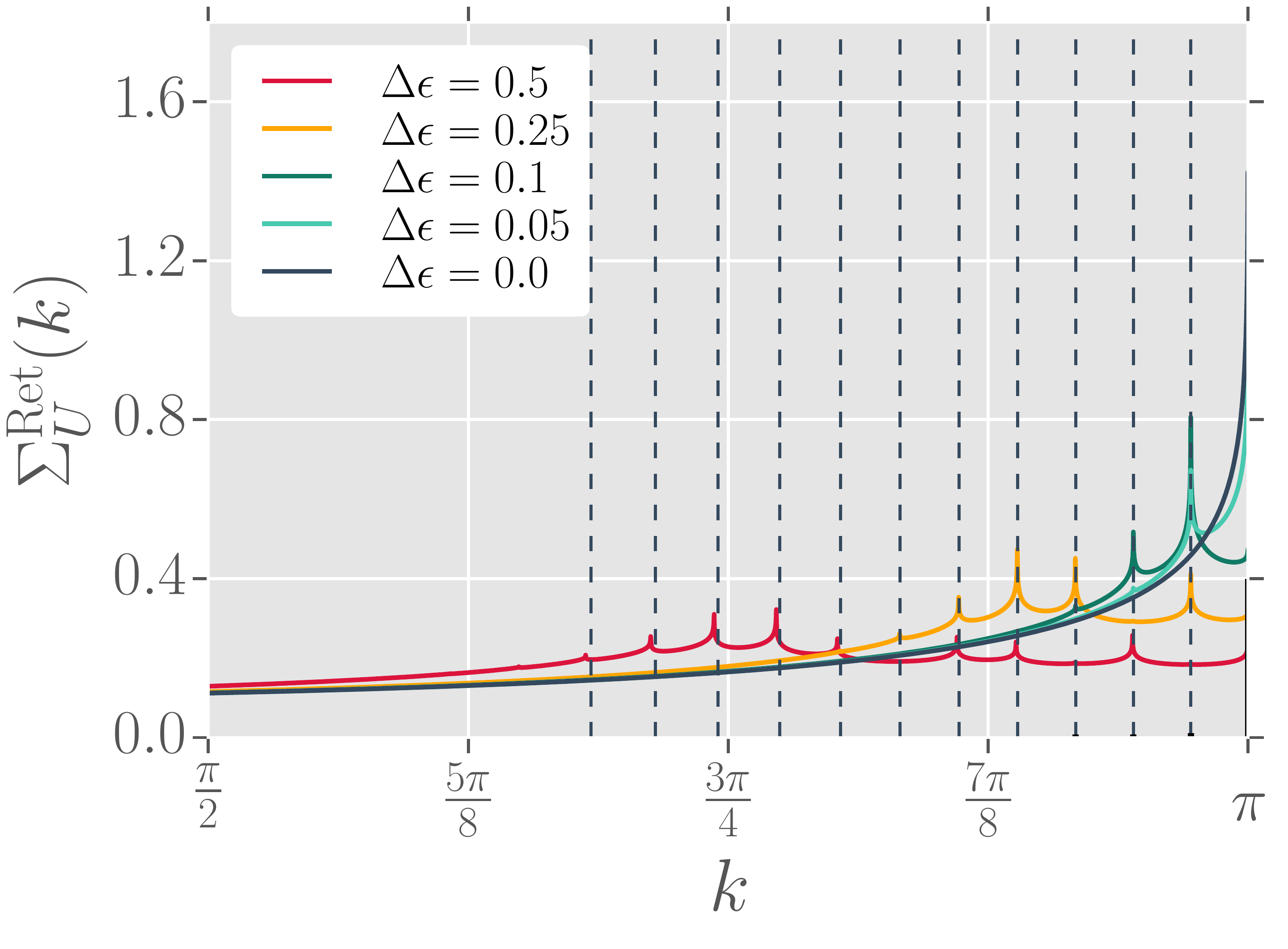}
\caption{Fourier transform of the renormalization of the effective hopping amplitudes  $\left.\Sigma^{\rm Ret}_U\right|_{0,i,0,i+1}$ (solid lines) as well as twice the wavevectors corresponding to energies at integer multiples of $\Omega$ (vertical dashed lines) for different $\Delta \epsilon$. The other paramters are $J_i=1$, $U=0.5$, $T=0$, $\Omega=0.1$, $\Gamma_i=0.001$ and $N=20000$.  }
\label{fig:Results3}
\end{figure}

To sum up, we find a superposition of approximately independent oscillations of the effective hoppings with period given by twice the wave numbers corresponding to energy $r\Omega$  and amplitude proportional to $\left|J_{r}\left(\frac{2\Delta\epsilon}{\Omega}\right)\right|^2$. In Fig.~\ref{fig:Results3} we show the Fourier transform of the self-energy $\left.\Sigma^{\rm Ret}_U\right|_{0,i,0,i+1}$ obtained by numerically solving the flow equations for different $\Delta \epsilon$ as solid colored lines. The position of twice the wavevectors corresponding to energy $r\Omega$ are indicated as dashed black lines. Peaks in the Fourier transform correspond to oscillations of $\left.\Sigma^{\rm Ret}_U\right|_{0,i,0,i+1}$ in real space. The relative heights of the different peaks in the Fourier transformation are consistent with the weighting factor $\left|J_{r}\left(\frac{2\Delta\epsilon}{\Omega}\right)\right|^2$ as expected, because the height relates to the amplitude of the oscillations. It is these oscillations in the effective hopping amplitudes that give rise to power-law suppressions of the boundary spectral function  at integer multiples of the external driving frequency as well as the sum of power-law suppressions in the conductance $G(T)$ discussed below.

\section{Results}
\label{sec:results}
\subsection{Boundary Spectral function}
\label{subsec:Spec}
The boundary (time-averaged) spectral function is related to the Green's functions 
\begin{align}
\rho_{\rm B}(\omega)&= -\frac{1}{\pi} {\rm Im}\left[ G^{\rm Ret}_{0,1,0,1}(\omega)\right]\notag\\
&= -\frac{1}{\pi} \sum_r \left|J_{r}\left(\frac{2\Delta\epsilon}{\Omega}\right)\right|^2 {\rm Im}\left[ \tilde G^{\rm Ret}_{1,1}(\omega+r\Omega)\right].\label{eq:sup_powerlaw}
\end{align}
After solving the flow equations \eqref{eq:SigmaU1} and \eqref{eq:SigmaU2} numerically we obtain the full interacting Green's function by including the self-energy generated by the FRG flow and thus the boundary spectral function in this manner. In this section we focus on constant  $\Gamma_i=\Gamma$ and $J_i=J$. 

\begin{figure}[t]
\centering
\includegraphics[width=\columnwidth]{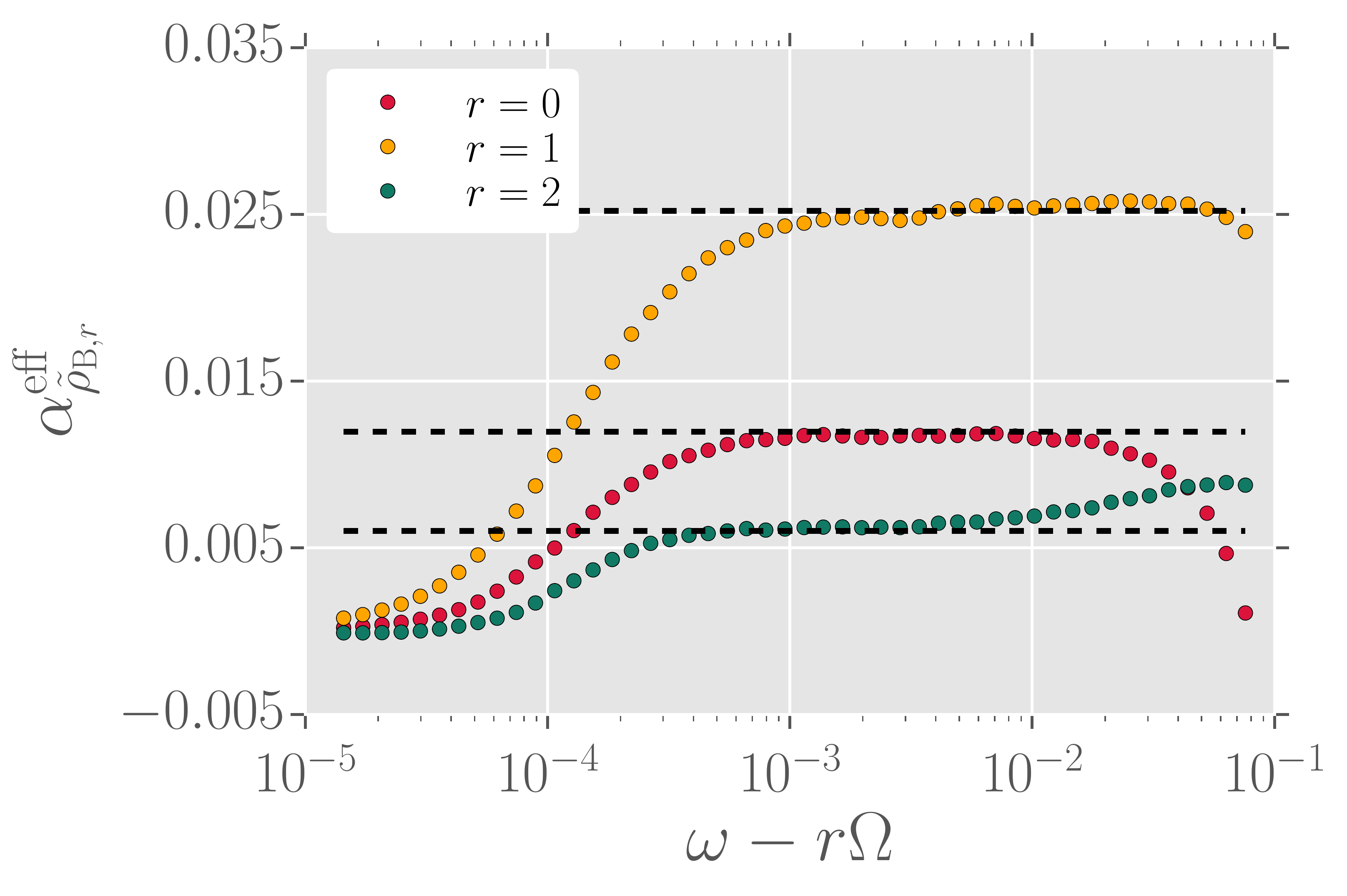}
\caption{Logarithmic derivative (effective exponent) of $\tilde \rho_{\rm B}$ (colored symbols) evaluated at different $\omega-i\Omega$. Dashed lines are the analytic prediction $\left|J_{r}\left(\frac{2\Delta\epsilon}{\Omega}\right)\right|^2 \alpha_{\rm B}$. The other parameters are $J_i=1$, $U=0.25$, $\Gamma_i=0.0001$, $\Delta \epsilon=0.17$, $\Omega=0.2$ and $N=10^6$.   }
\label{fig:Results4}
\end{figure}

We concentrate first on the low frequency $|\omega| \to 0$ behavior and note that it is a superposition of the behavior of 
 \begin{equation}
\tilde \rho_{\rm B}=-\frac{1}{\pi}{\rm Im}\left[\tilde G_{11}^{\rm Ret}(\omega)\right] 
\end{equation}
evaluated at frequencies $|\omega+r\Omega| \to 0$.
To make analytic progress we neglect the order $U^2$ contribution of the mutual  influence of the oscillations in the effective hoppings generated by the flow (see above). 
Then $\tilde \rho_{\rm B}$ exhibits a power-law suppression at each value of $|\omega+r\Omega| \to 0$ due to the renormalization group procedure as described above.  The exponents are given by
\begin{equation}
 \alpha_{{\rm B},r}=\left|J_{r}\left(\frac{2\Delta\epsilon}{\Omega}\right)\right|^2 \alpha_{{\rm B}}.
 \label{eq:alphabr}
 \end{equation} 
In the following we will frequently  utilize the log-derivative (realized as centered differences for numerical data)
\begin{equation}
\alpha_{y(x)}=\frac{d \log(y)}{d \log(x)},
\end{equation} 
of a function $y(x)$, which becomes a constant of value $\alpha$ if $y=x^\alpha$ and is thus a very sensitive measure to reveal power-law behavior unambiguously.
The log-derivative of $\tilde \rho_{\rm B}$ at frequencies $|\omega+r\Omega| \to 0$ is shown in Fig.~\ref{fig:Results4}. We compare the analytic results of Eq.~\eqref{eq:alphabr} to the full numerics and find excellent agreement, up to deviations at low and high frequencies, where the universal power-law behavior is cut off (at low frequencies by $\Gamma$ and at high frequencies by the bandwidth).

Equation~\eqref{eq:sup_powerlaw} implies that the observable $\rho_{\rm B}$ is an additive superposition of $\tilde \rho_{\rm B}$ and thus of power-law behavior, which yields the form given in the introduction Eq.~\eqref{eq:sup_powerlaw_intro} for $\rho_{\rm B}(\omega)$.
The different contributions of the sum have prefactors  $\left|J_{r}\left(\frac{2\Delta\epsilon}{\Omega}\right)\right|^2 m_r$, where $m_r$ depends on the microscopic details of the model and to zeroth approximation is given by the value of the $U=0$, non-interacting Green's function $m_r={\rm Im}\left[\tilde  G^{{\rm Ret},U=0}_{i=1,i'=1}(\omega+r\Omega)\right]$. 
 The first factor $\left|J_{r}\left(\frac{2\Delta\epsilon}{\Omega}\right)\right|$ defines a value of $r_c$ at which the sum can be cut off as contributions outside of $-r_c(2\Delta\epsilon/\Omega)\leq r\leq r_c(2\Delta\epsilon/\Omega)$ become negligible. 
 
\begin{figure}[t]
\centering
\includegraphics[width=\columnwidth]{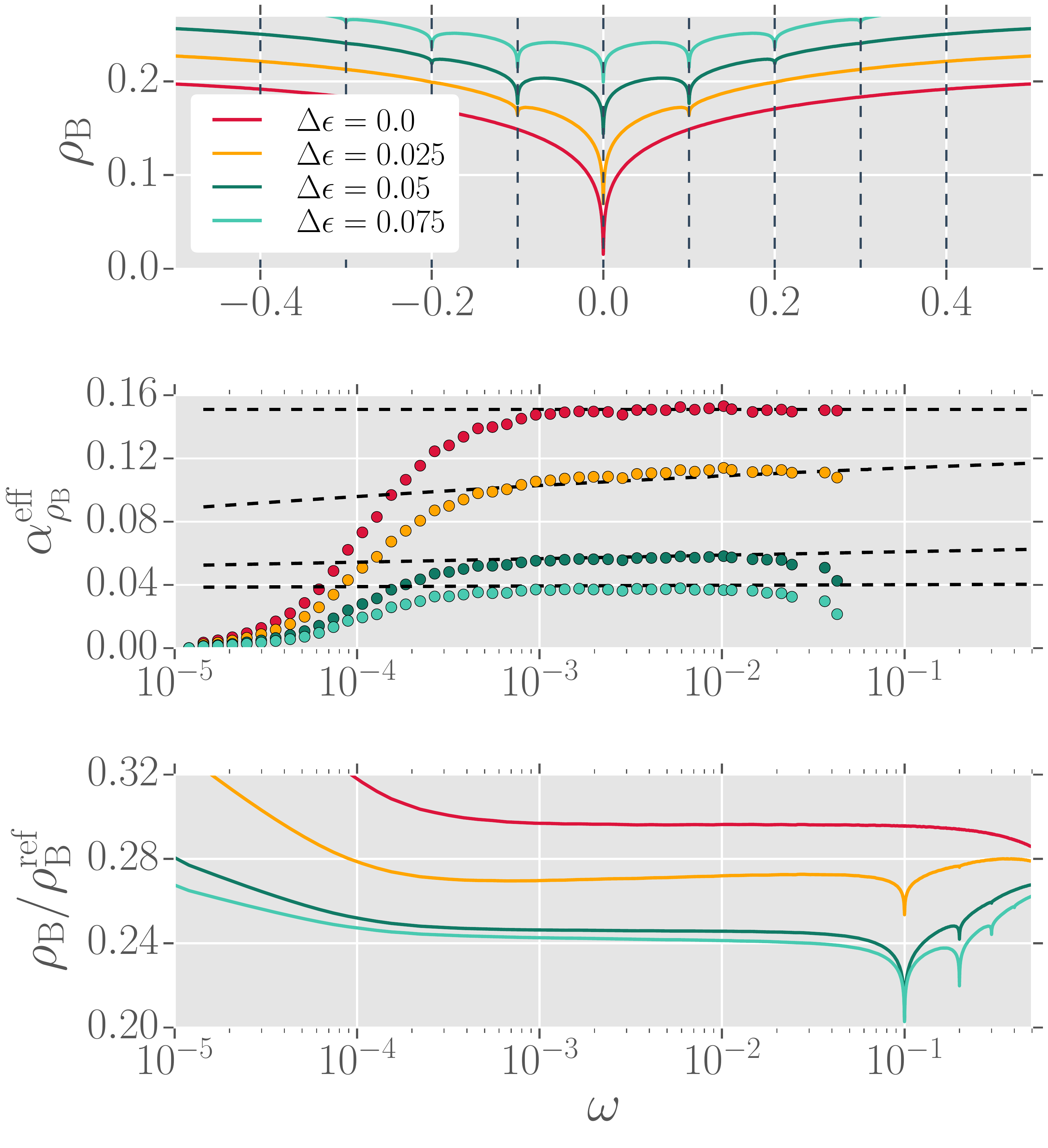}
\caption{Top panel: Spectral function of a quantum wire (solid colored lines), where the onsite energies are varied periodically in time with different driving amplitudes $\Delta\epsilon$. The parameters are $J_i=1$, $U=0.5$, $\Gamma_i=0.0001$, $\Omega=0.5$, $N=10^5$ and different colors indicate different amplitudes of driving $\Delta\epsilon$. For clarity of depiction the lines are shifted vertically with respect to each other. The power-law suppression of the boundary spectral function at $\omega=0$ in equilibrium ($\Delta\epsilon=0$) is split into multiple suppressions at $\omega=k\Omega$ (indicated by dashed black lines) when the periodic driving is turned on. Middle panel: the logarithmic derivative of the spectral function to analyze the additive superposition of power-laws. For a single power-law the logarithmic derivative would be constant. The additive superposition leads to an almost straight line, but with finite slope indicating that more than power-law are relevant to the sum. The dashed black line is the prediction Eq.~\eqref{eq:sup_powerlaw_intro} for $r=0$ and with $c_{0,r'}=1$. The finite size of $N=10^5$ lattice sites as well as finite $\Gamma_i$ introduce  cut-offs to the suppression of the boundary spectral function (as in equilibrium) proportional to the maximum of the level splitting $~1/N$ and $\Gamma_i$. Bottom panel the ratio between the spectral function and the prediction Eq.~\eqref{eq:sup_powerlaw_intro}. Clearly in the regime of small $\omega$, but before the cutoff scale is reached, the ratio becomes flat, validating the analytic finding Eq.~\eqref{eq:sup_powerlaw_intro}.  }
\label{fig:Results1}
\end{figure}

In the limit of large bandwidth, meaning that $4J\gg r_c(2\Delta\epsilon/\Omega) \Omega$, the factor $m_r\equiv c_{0,r}$ is approximately constant and the analytic prediction Eq.~\eqref{eq:sup_powerlaw_intro} can be tested against the numerics (beyond this limit one needs to take into account the band curvature and predicting $c_{0,r}$ becomes more involved). Representative results are summarized in the upper panel of Fig.~\ref{fig:Results1}, where we show numerically determined  boundary spectral functions. The single power-law suppression found in the equilibrium  case ($\Delta\epsilon=0$) is split into a series of suppressions at $r\Omega$ indicated by vertical dashed lines. In the middle panel of Fig.~\ref{fig:Results1} we compare the logarithmic derivative of the numerical data to the analytical prediction and find excellent agreement. The lines do not approach a constant value at small frequencies, which emphasizes that the full functional form of $\rho_{\rm B}(\omega)$ is not a (single) power-law around $\omega=r\Omega$. This holds down to frequencies as small as the cut-off provided by the system-size or $\Gamma_i$, whichever is larger. In the bottom panel of Fig.~\ref{fig:Results1} we compare the ratio between numerics and the analutic prediction Eq.~\eqref{eq:sup_powerlaw_intro} to find perfect agreement.
 
So far we have concentrated on the $\omega\to 0$ behavior of $\rho_{\rm B}$, analogous ideas can be used to study the  boundary spectral function at all other $|\omega-r\Omega| \to 0$ as well leading to the prediction of Eq.~\eqref{eq:sup_powerlaw_intro} for these frequencies.

\begin{figure}[t]
\centering
\includegraphics[width=\columnwidth]{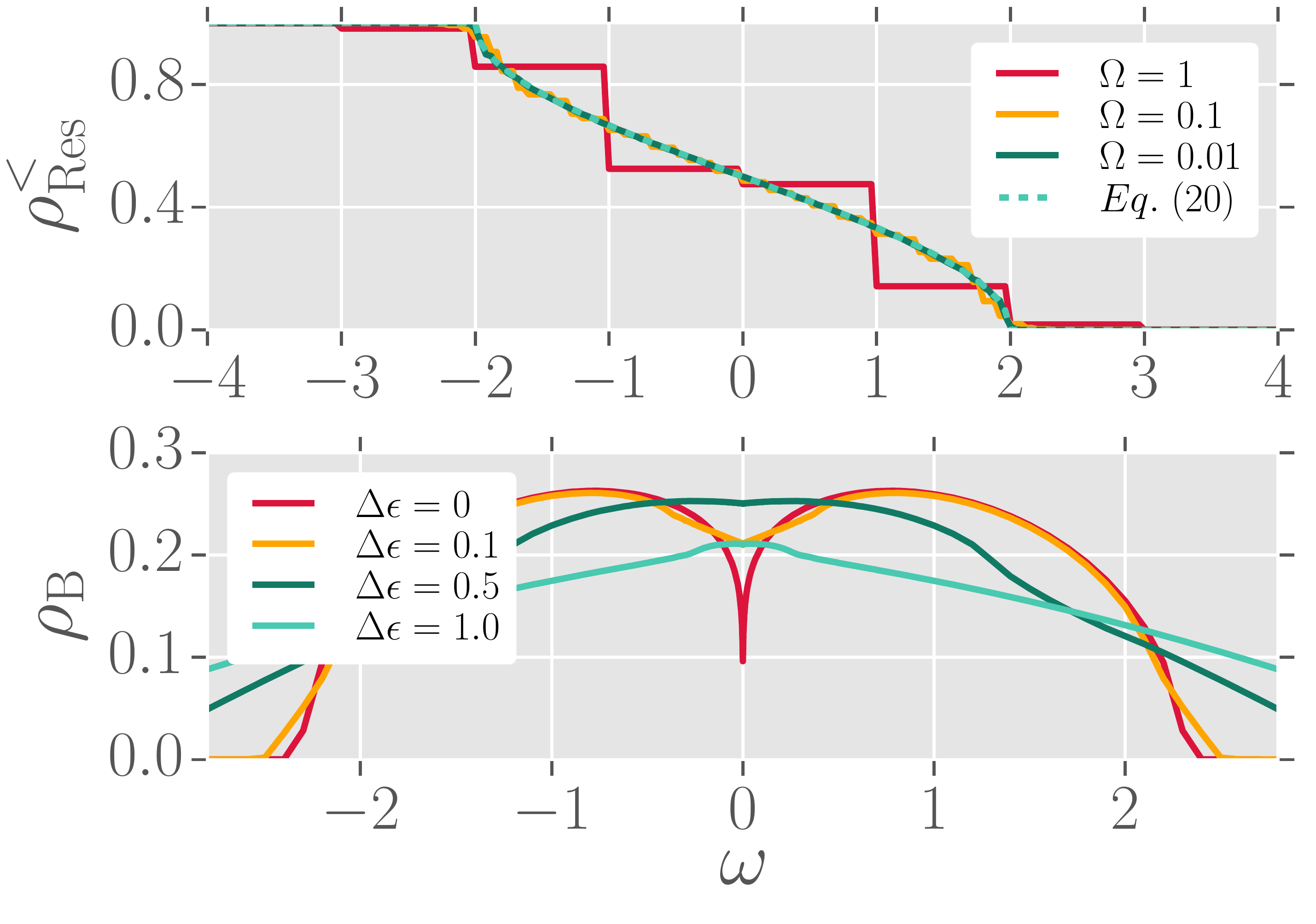}
\caption{ Upper Panel: Joint reservoir distribution function Eq.~\eqref{eq:jrd} for $\Delta\epsilon=1$ and different $\Omega$. For $\Omega\to0$ we recover Eq.~\eqref{eq:asym_jrd}
Lower Panel: Boundary spectral function of a very slowly driven quantum wire with $\Omega=0.01$. The other parameters are as in Fig.~\ref{fig:Results1}.}
\label{fig:Results2}
\end{figure}

Up to this point we focused on $\Omega\ll 4J$, in the opposite, anti-adiabatic, limit of $\Omega\gg 4J_i$ the oscillations are very fast and, trivially, the known undriven equilibrium result is recovered. Another interesting limit is to approach   $\Omega\to 0$ and keeping $\Delta\epsilon$ fixed, for which we find a dimensionality crossover to an effectively two dimensional system. This can be understood from the effective Floquet Hamiltonian. As $r_c(2\Delta\epsilon/\Omega)\to \infty$ all higher Floquet modes become relevant. We define the joint reservoir distribution function as 
\begin{align}
\rho_{\rm Res}^<(\omega)=&\sum\limits_{n=-\infty}^\infty J_n(2\Delta \epsilon/\Omega)^2\Theta(n\Omega-\omega ),\label{eq:jrd}
\\
\stackrel{\Omega\to0}{=}&\;\;\;\;{\rm Re}\left[\frac{\arccos\left(\frac{\omega}{2\Delta\epsilon}\right)}{\pi}\right],\label{eq:asym_jrd}
\end{align}
which becomes smooth and the power-law suppression (arising from the sharp jump at the Fermi-edge in equilibrium) vanishes. In this limit the infinite sum of power-laws sums up to a regular contribution, which does not require an RG treatment. The upper panel of Fig.~\ref{fig:Results2} illustrates how the joint reservoir distribution function evolves from a multiple step function to a continuous form as $\Omega$ approaches $0$. The lower panel of    Fig.~\ref{fig:Results2} shows the boundary spectral function for slow driving $\Omega=0.01$. As $\Delta\epsilon$ becomes comparable to $J$  indeed an  effectively two dimensional system is realized and the power-law suppressions are wiped out.

\subsection{Conductance}
\label{sec:conc}
\label{subsec:Cond}
A more directly accessible, transport quantity is the (time-averaged) linear conductance
\begin{equation}
G(T)=\frac{\Omega}{2\pi}\int\limits_{0}^{\frac{\Omega}{2\pi}}\frac{\partial J_{\rm L}(t)}{\partial V}dt
\end{equation}
where $J_{\rm L}$ is the current leaving (e.g.) the left reservoir. In this section, to establish a transport setup, we concentrate on the case of the wire being end-contacted by two reservoirs $\Gamma_1=\Gamma_N=\Gamma$ and $\Gamma_{1<i<N}=0$. In our setup it turns out that (because the occupancies do not change in time, see above)
\begin{equation}
J_{\rm L}(t)=-J_{\rm R}(t)
\end{equation} 
and thus the definition of the conductance is unambiguous. 
The Fourier components of the current and with it of the time -averaged conductance can be calculated by\cite{Eissing16a,Eissing16b} 
\begin{align}
G(T)=\frac{\partial}{\partial V} \frac{1}{2\pi} \int d\omega&\;\bigg[ \Sigma^{\rm Ret, L}_{0,1,0,1} G^<_{0,1,0,1}(\omega)\notag\\&-  G^{\rm Ret}_{0,1,0,1}(\omega) \Sigma^{<, L}_{0,1,0,1}\bigg],
\end{align}
where $\Sigma^{L}$ includes only the contribution of the self-energy induced by the left reservoir. For the considered driving one can recast the conductance into the Landauer-B\"uttiker form (within our lowest order truncated flow equations)
\begin{align}
G(T)=&\sum\limits_{r=-\infty}^\infty G^{r}(T)\\
G^{r}(T)&=\Gamma^2  \left|J_{r}\left(\frac{2\Delta\epsilon}{\Omega}\right)\right|^2  \int d\omega\; f'(\omega+r\Omega) \left|\tilde G^{\rm Ret}_{1,N}(\omega)\right|^2.
\end{align}

\begin{figure}[t]
\centering
\includegraphics[width=\columnwidth]{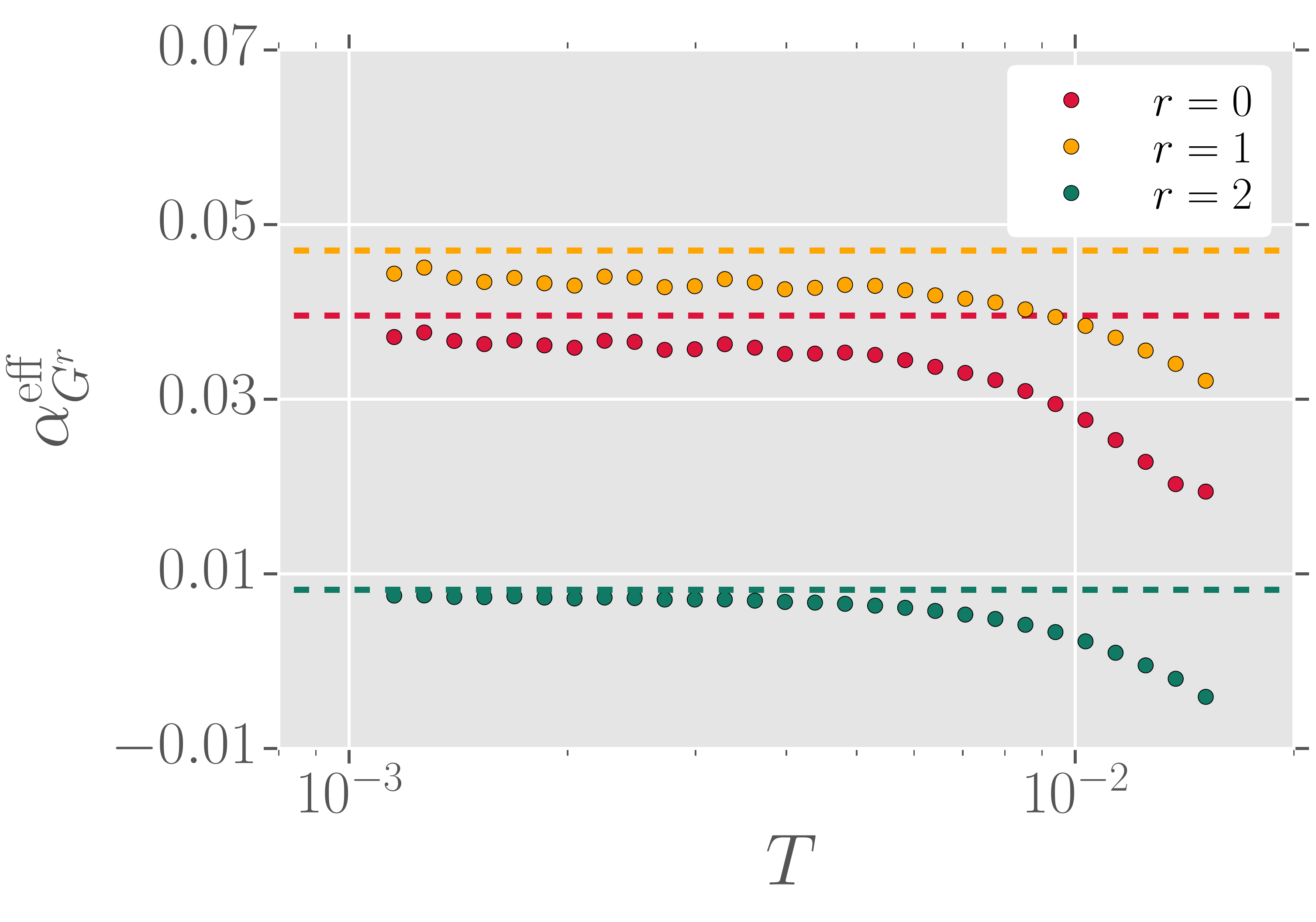}
\caption{Logarithmic derivative (effective exponent) of the different constituents $G^i(T)$ contributing to the sum of power-laws in the suppression of the conductance in dependence of temperature $T$. The numerical data (symbols) agrees upto $\mathcal{O}(U^2)$ with the predicted exponents $\left|J_{r}\left(\frac{2\Delta\epsilon}{\Omega}\right)\right|^2 \alpha_{{\rm B}}$ (dashed lines) of Eq.~\eqref{Eq:Gana1}. The other parameters are  $J=1$, $U=0.5$, $\Omega=0.1$, $\Delta\epsilon=0.075$, $\Gamma=0.01$ and $N=10^5$. }
\label{fig:Results5}
\end{figure}

\subsubsection{Impurities at contacts}
First we consider an impurity free wire $J_i=J$ and $U_i=U$. As the interaction is abruptly switched off at the boundaries, where the wire is end-contacted to non-interacting reservoirs, this creates two impurities at both of these ends for which fermions tunnel from Fermi- to Luttinger liquid or vice versa. The same calculation as performed above for the  boundary spectral function adepted to the conductance leads to the analytic prediction (where again we disregard the order $U^2$ contribution of the mutual influence of the oscillations in the effective hoppings generated by the flow to make analytic progress)
\begin{align}
G(T)\sim &\sum\limits_{r=-\infty}^\infty G^{r}(T)\\
G^{r}(T)&=\left|J_{r}\left(\frac{2\Delta\epsilon}{\Omega}\right)\right|^2 \left(\frac{T}{T_0}\right)^{ \left|J_{r}\left(\frac{2\Delta\epsilon}{\Omega}\right)\right|^2 \alpha_B}.\label{Eq:Gana1}
\end{align}
We find a similar behavior as for the boundary spectral function but now in the temperature scaling of the conductance through the wire: the single power-law with exponent $\alpha_{\rm B}$ found in the equilibrium, undriven case is replaced by an infinite sum of power-laws governed by impurity independent exponents given in Eq.~\eqref{eq:alphabr}. This analytic prediction is checked for representative parameters in Fig.~\ref{fig:Results5} against the full numerics solving Eqs.~\eqref{eq:SigmaU1} and \eqref{eq:SigmaU2}.

\subsubsection{Single impurity at center}

Next we study transport through a single impurity at the center of the chain $J_{i\neq N/2}=J$ and $J_{i= N/2}=J'$. We want to model a homogeneous system and thus need to match the band of the reservoirs to that of the wire without the impurity. We can do this efficiently within the framework discussed so far (which treated reservoirs in the wide band limit only) by  restricting the interaction in the wire to a central and much smaller region  in space around the impurity of size $N_{\rm int}$ then the full length of the wire $N\gg N_{\rm int} $. Additionally, we need to avoid scattering from the end-contacts of the interacting and non-interacting regions. We thus smoothly turn on (and off) the interactions at the two boundaries of the region given by $N_{\rm int}$  according to $\arctan(i/2)$ over a width of $N_{\rm int}/10$ lattice sites to a constant value of $U$, while outside of the region given by the $N_{\rm int}$ lattice sites around the impurity $U=0$. It  was shown\cite{Meden03,Enss05,Karrasch15} that turning the interactions on and off according to this $\arctan$ form leads to negligible scattering from the contacts in equilibrium and thus the transport physics through a single impurity (as introduced in the center of the interacting wire) is probed.

This setup models the transport through an interacting wire with a weakened link, which in equilibrium was found to reveal an interesting effect of strong correlations:\cite{Kane92,Meden03,Enss05} an arbitrary small impurity (reduction in hopping amplitude) cuts the wire in two at zero temperature and for infinite wire length, suppressing transport completely. At non-zero temperature (or wire length) this suppression is incomplete and follows a power-law form with twice the boundary exponent $G\sim T^{2\alpha_{\rm B}}$.  In this context FRG was instrumental to unravel the existence of a single-parameter scaling function that connects the weak to strong impurity flow of the conductance (to zero) without an intermediate fix point as the temperature is lowered.\cite{Meden03,Enss05}

\begin{figure}[t]
\centering
\includegraphics[width=\columnwidth]{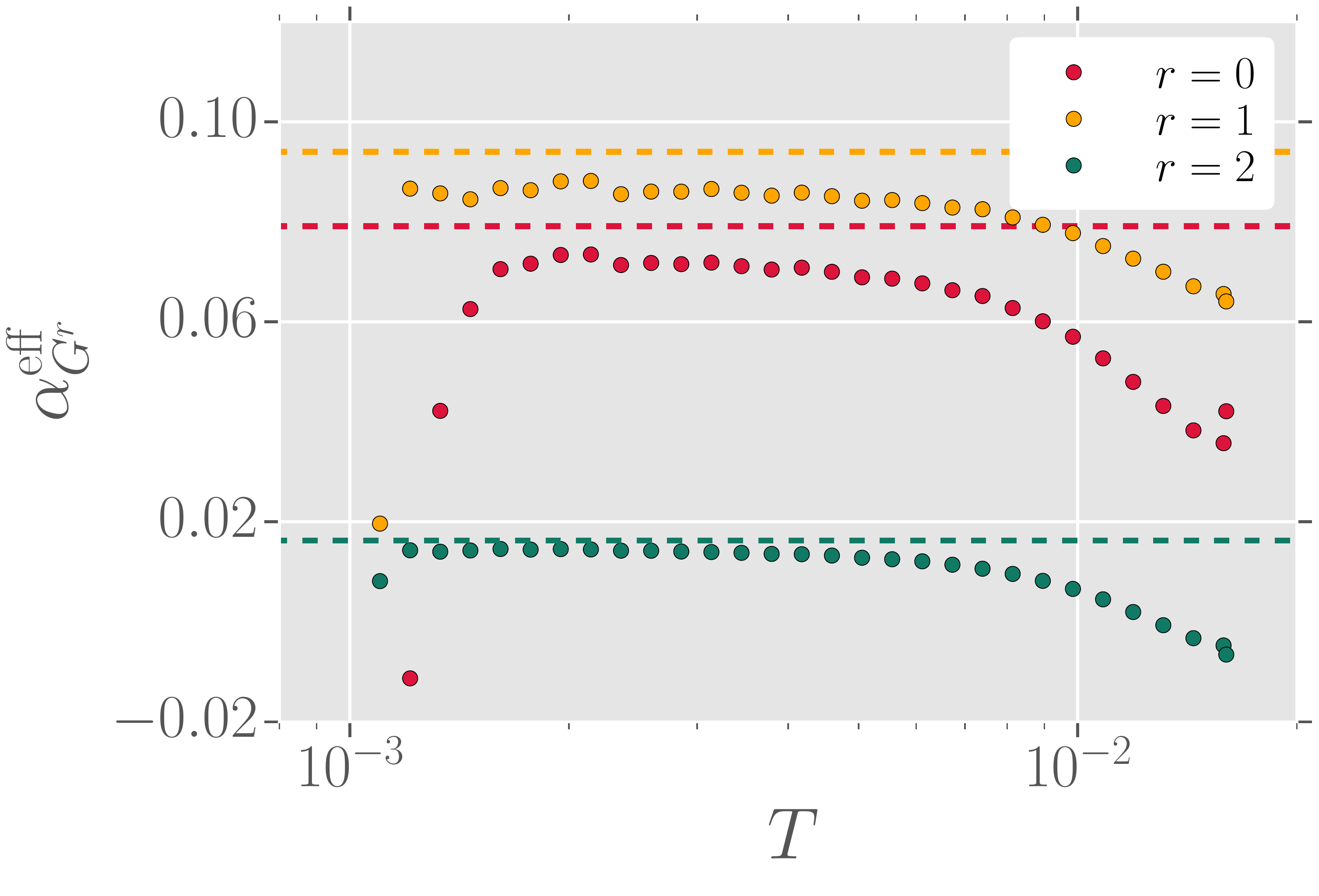}
\caption{ Logarithmic derivative (effective exponent) of the different constituents $G^i(T)$ contributing to the sum of power-laws in the suppression of the conductance in dependence of temperature $T$. The numerical data (symbols) agrees upto $\mathcal{O}(U^2)$ with the predicted exponents $2\left|J_{r}\left(\frac{2\Delta\epsilon}{\Omega}\right)\right|^2 \alpha_{{\rm B}}$ (dashed lines) of Eq.~\eqref{Eq:Gana2}. The other parameters are  $J=1$, $J'=0.1$, $U=0.5$, $\Omega=0.1$, $\Delta\epsilon=0.075$, $\Gamma=0.01$, $N=10^7$ and $N_{\rm int}=10^5$. }
\label{fig:Results6}
\end{figure}

Similar to the analysis of the boundary spectral function, if we neglect band curvature over the ranges of energy we are interested in, we find
\begin{align}
G(T)\sim &\sum\limits_{r=-\infty}^\infty G^{r}(T)\\
G^{r}(T)&=\left|J_{r}\left(\frac{2\Delta\epsilon}{\Omega}\right)\right|^2 \left(\frac{T}{T_0}\right)^{2 \left|J_{r}\left(\frac{2\Delta\epsilon}{\Omega}\right)\right|^2 \alpha_B}\label{Eq:Gana2}
\end{align}
for strong impurities $J'\ll J$ from our FRG analysis. Again, the single power-law is turned into an infinite sum of power-laws with impurity independent exponents now given by twice the value of Eq.~\eqref{eq:alphabr}. This analytic prediction is checked for representative parameters in Fig.~\ref{fig:Results6}.

\begin{figure}[t]
\centering
\includegraphics[width=\columnwidth]{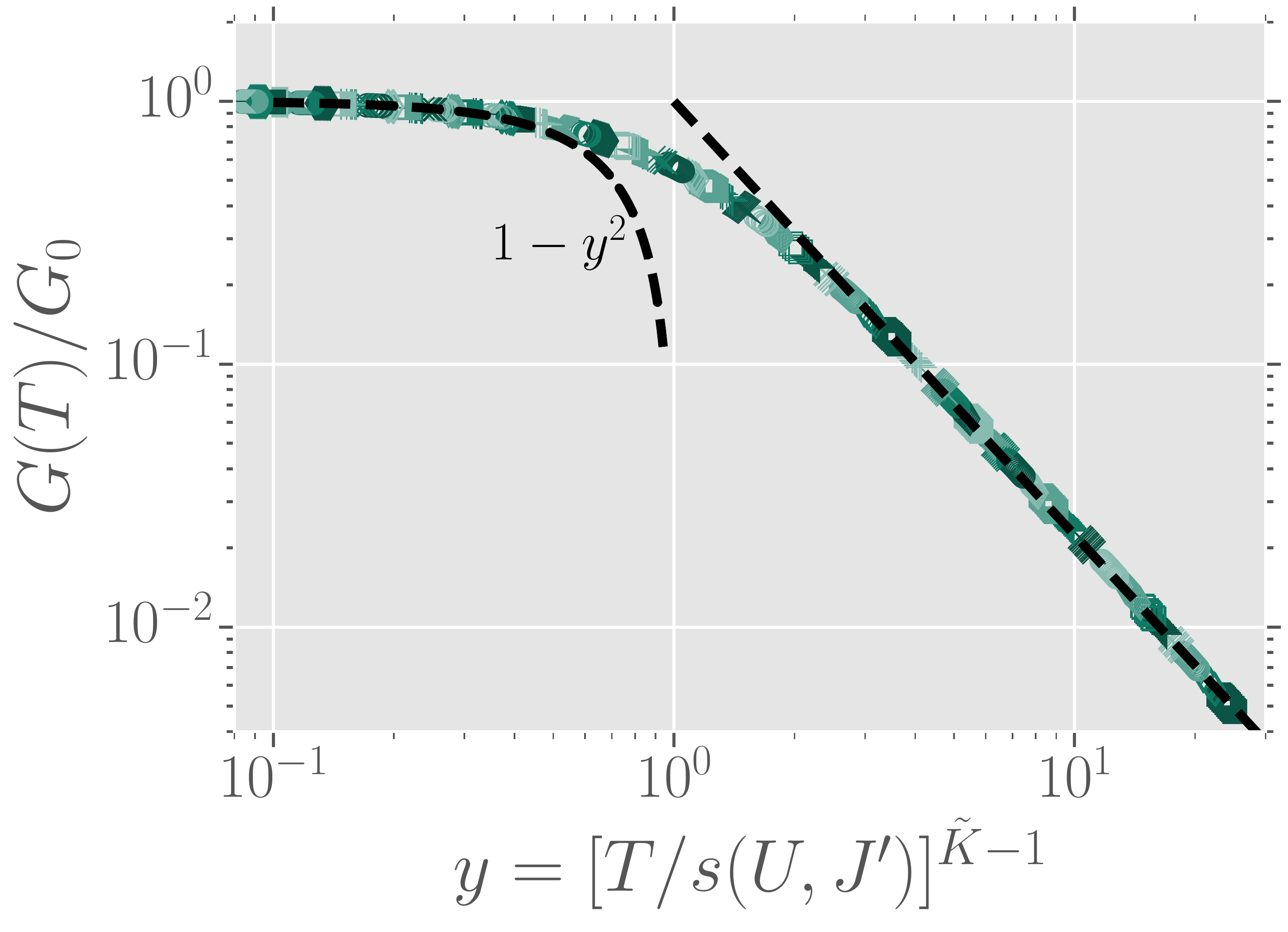}
\caption{ One-parameter scaling plot of $G^0(T)$. Different symbols and colors represent results obtained for different impurity strengths $J'$. The dashed lines indicate the expected behavior ar small and large $y$. The other parameters are the same as in Fig.~\ref{fig:Results6}.  }
\label{fig:Results7}
\end{figure}

We finally show that in the steady state of the periodically driven lattice model there exists no fixed point in between the flow of a weak and a strong impurity. We will do so on the level of the constituents $G^r(T)$, which enter additively in $G(T)$. To achieve this we calculate $G(T)$ for many values of $J'$ at constant $U$. In equilibrium the  one-parameter scaling ansatz $G^r(T)$ = $\tilde G^r(y)$ where $y = (T /s)^{\tilde K -1}$ and $s(U,J')$ is a  nonuniversal scale can be used to connect the flow of weak to strong impurities. We utilize an analogous scaling form, but with $\tilde K$ to be the Luttinger liquid parameter with respect to the reduced interaction $\tilde U=\left|J_{r}\left(\frac{2\Delta\epsilon}{\Omega}\right)\right|^2 U$.  
In doing so, all data can be collapsed on a single curve continuously
connecting the weak ($y \to 0$) and strong ($y \to \infty$) impurity fixed points. This is exemplified for $G^{0}$ in Fig.~\ref{fig:Results7}. In our FRG treatment we thus find that every component contributing to the sum describing the conductance behaves similarly to the impurity physics of equilibrium.\cite{Kane92}

\section{Conclusions}
We analyzed the long time behavior of periodically driven quantum wires. We found that there is a crossover controlled by the ratio $\Delta\epsilon/\Omega$ of the driving strength $\Delta\epsilon$ and the frequency of driving $\Omega$ from a modified Luttinger liquid picture to an effectively higher dimensional system. We concentrated on the boundary spectral function and the conductance as  hallmark observables indicating Luttinger liquid behavior in equilibrium. We showed analytically that in the modified Luttinger liquid regime the single power-law suppression  of the boundary spectral function at the Fermi-level with interaction dependent exponent is split into multiple suppressions at integer multiples of the external drive frequency $\Omega$. Each suppressions turns from a single power-law to an infinite sum of power-laws with exponents being controlled by the drive. In the opposite regime where the system behaves as an effectively higher dimensional one we find that the power-law is wiped out as expected as the Luttinger liquid paradigm seizes to hold in higher dimensions. Similar consequences can be derived for the conductance where again the single power-law suppression of transport in dependency of temperature is replaced by an infinite sum of power-laws, where each constituent can be understood analytically. We report that, similar to equilibrium,\cite{Kane92} the weak and strong impurity flow are continuously connected by a single parameter scaling function, without an intermediate fix point showing up. A similar finding was reported after a quench of a Luttinger liquid within the same FRG treatment.\cite{Kennes13} It was argued\cite{Mitra15,Mitra17} that within a perturbative RG\cite{Mitra11,Mitra12,Mitra13} including higher order terms might introduce changes to this behavior at higher orders in $U$. Conducting such a study for the periodically driven case would be highly interesting.  

Another intriguing avenue of future research could be to tackle more complicated driving protocols. Here, we focused on a particularly simple form of driving for which all the onsite energies are varied periodically in time. This allowed us to obtain fully analytical results. Studying a more complicated driving protocol should still be numerically  feasible within the approach we develop here. An interesting question along these lines would be to study the influence of AC bias voltages or pumping  with light field. Furthermore, extending the study presented here for spinless fermions to the Hubbard model would be of interest. 

\begin{acknowledgements}
We thank Severin Jakobs and Volker Meden for many helpful comments.
This work was supported by the Basic Energy Sciences Program of the U. S. Department of Energy under Grant No. SC-0012375 and by DFG KE 2115/1-1.
Simulations were performed with computing resources granted by RWTH Aachen University under project rwth0013.
We also acknowledge the hospitality of the Center for Computational  Quantum Physics  of  the  Flatiron  Institute.
\end{acknowledgements}

{}


\begin{thebibliography}{}


\bibitem{Bloch08}
  I.~Bloch, J.~Dalibard, and W.~Zwerger, Rev.~Mod.~Phys. {\bf 80}, 885 (2008).


\bibitem{Yu91}
G. Yu, C. H. Lee, A. J. Heeger, N. Herron, and E. M. McCarron 
  Phys. Rev. Lett. {\bf 67}, 2581 (1991).

\bibitem{Miyano97}
  K. Miyano, T. Tanaka, Y. Tomioka, and Y. Tokura,
  Phys. Rev. Lett. {\bf 78}, 4257 (1997).

\bibitem{Wall11}
  S. Wall,	D. Brida,	S. R. Clark,	H. P. Ehrke,	D. Jaksch,	A. Ardavan,	S. Bonora,	H. Uemura,	Y. Takahashi,	T. Hasegawa,	H. Okamoto,	G. Cerullo	and A. Cavalleri,
  Nature Physics {\bf 7}, 114 (2011).


\bibitem{Blumenstein11}
  C. Blumenstein, J. Sch\"afer, S. Mietke, S. Meyer, A. Dollinger, M. Lochner, 
  X.Y. Cui, L. Patthey, R. Matzdorf, and R. Claessen,
  Nature Physics {\bf 7}, 776 (2011).

\bibitem{Lignier07}
H. Lignier, et. al.,
Phys. Rev. Lett. {\bf 99}, 220403 (2007).

\bibitem{Kierig08}
E. Kierig, U. Schnorrberger, A. Schietinger, J. Tomkovic, M.K. Oberthaler,
Phys. Rev. Lett. {\bf 100}, 190405 (2008).

\bibitem{Sias08}
C. Sias, et. al.,
Phys. Rev. Lett. {\bf 100}, 040404 (2008).



\bibitem{Wilczek}
F. Wilczek,
Phys. Rev. Lett. {\bf 109}, 160401 (2012)

\bibitem{Else16}
D. V. Else, B. Bauer and C. Nayak,
Phys. Rev. Lett. {\bf 117}, 090402 (2016).

\bibitem{Khemani16a}
V. Khemani , A. Lazarides, R. Moessner and S. L. Sondhi,
Phys. Rev. Lett. {\bf 116}, 250401 (2016).

\bibitem{Keyserlingk16}
C. W. von Keyserlingk, Vedika Khemani and S. L. Sondhi,
Phys. Rev. B {\bf 94}, 085112 (2016).

\bibitem{Khemani16b}
Vedika Khemani, C. W. von Keyserlingk and S. L. Sondhi,
arXiv:1612.08758.

\bibitem{Moessner17}
R. Moessner	and S. L. Sondhi,
Nature Physics {\bf 13}, 424 (2017).


\bibitem{Bordia17}
P. Bordia, et. al.,
Nat. Phys. {\bf 13}, 466 (2017).

\bibitem{Zhang17}
J. Zhang, et. al.,
Nature {\bf 543}, 217 (2017).

\bibitem{Choi17}
S. Choi, et. al., system.
Nature {\bf 543}, 221 (2017).

\bibitem{Else17}
D. V. Else, B. Bauer, and C. Nayak
Phys. Rev. X 7,  011026 (2017).

\bibitem{FTI}
T. Oka and H. Aoki,
Phys. Rev. B {\bf 79}, 081406 (2009);
J. I. Inoue and A. Tanaka,
Phys. Rev. Lett. {\bf  105}, 017401 (2010);
T. Kitagawa, E. Berg, M. Rudner, and E. Demler, Phys. Rev. B {\bf 82}, 235114 (2010);
N. H. Lindner, G. Refael, and V. Galitski,
Nat. Phys. {\bf 7}, 490 (2011);
T. Kitagawa, T. Oka, A. Brataas, L. Fu, and E. Demler,
Phys. Rev. B {\bf 84}, 235108 (2011);
N. H. Lindner, D. L. Bergman, G. Refael, and V. Galitski,
Phys. Rev. B {\bf 87}, 235131 (2013);
A. G\'omez-Le\'on and G. Platero,
Phys. Rev. Lett. {\bf 110}, 200403 (2013);
Y. T. Katan and D. Podolsky,
Phys. Rev. Lett. {\bf 110}, 016802 (2013);
P.M. Perez-Piskunow, G.Usaj, C.A. Balseiro and L.E.F. FoaTorres,
Phys. Rev. B {\bf 89}, 121401(R) (2014).

\bibitem{QDp}
M. Switkes, C. M. Marcus, K. Campman, and A. C.
Gossard, Science {\bf 283}, 1905 (1999);
B. Roche, R. P. Riwar, B. Voisin, E. Dupont-Ferrier, R.
Wacquez, M. Vinet, M. Sanquer, J. Splettstoesser, and X.
Jehl, Nat. Commun. {\bf 4}, 1581 (2013);
H. T. M. Nghiem and T. A. Costi
Phys. Rev. B {\bf 90}, 035129 (2014);
M. M. Odashima and C. H. Lewenkopf
Phys. Rev. B {\bf 95}, 104301 (2017);
T. J. Suzuki
Phys. Rev. B {\bf 95}, 241302 (2017);
S. A. Reyes, D. Thuberg, D. P\'erez,  C. Dauer, S. Eggert, 2017 New J. Phys. {\bf 19} 043029;
D. Thuberg, E. Munoz, S. Eggert, S. A. Reyes,
Phys. Rev. Lett. {\bf 119}, 267701 (2017).


\bibitem{Eissing16a}
A. K. Eissing, V. Meden, and D.M. Kennes,
Phys. Rev. Lett. {\bf 116}, 026801 (2016).
\bibitem{Eissing16b}
A. K. Eissing, V. Meden, and D.M. Kennes,
Phys. Rev. B {\bf 94}, 245116 (2016).

\bibitem{LLperio}
Yu. Kagan and L. A. Manakova
Phys. Rev. A {\bf 80}, 023625 (2009);
C. D. Graf, G. Weick, and E. Mariani,
EPL {\bf 89}, 40005 (2010);
D. Poletti and C. Kollath,
Phys. Rev. A {\bf 84}, 013615 (2011);
S. Pielawa
Phys. Rev. A {\bf 83}, 013628;
M. Bukov and M. Heyl,
Phys. Rev. B {\bf 86}, 054304 (2012).


\bibitem{Metzner12}
  W.~Metzner, M.~Salmhofer, C.~Honerkamp, V.~Meden, and K.~Sch\"onhammer,
  Rev.~Mod.~Phys.~{\bf 84}, 299 (2012). 
  
\bibitem{Kopietz10}
P. Kopietz, L. Bartosch, and F. Sch\"utz, Introduction to
the Functional Renormalization Group, Lecture Notes in
Physics (Springer-Verlag Berlin Heidelberg, 2010).


\bibitem{Meden02} 
V. Meden, W. Metzner, U. Schollw\"ock, and K. Sch\"onhammer
J. of Low Temp. Physics {\bf 126}, 1147 (2002) 

\bibitem{Andergassen04} 
  S.~Andergassen, T.~Enss, V.~Meden, W.~Metzner, U.~Schollw\"ock, and K.~Sch\"onhammer,
  Phys.~Rev.~B {\bf 70}, 075102 (2004).

\bibitem{Enss05}
  T.~Enss, V.~Meden, S.~Andergassen, X.~Barnabe-Theriault, W.~Metzner, and K.~Sch\"onhammer,
  Phys.~Rev.~B {\bf 71}, 155401 (2005).
 
\bibitem{Jakobs07}
S.G. Jakobs, V. Meden, and H. Schoeller
Phys. Rev. Lett. 99, 150603 (2007).

\bibitem{Kennes13}
  D. M. Kennes and V. Meden,
  Phys. Rev. B {\bf 88}, 165131 (2013).
  
\bibitem{Hacker70}
K. Hacker and G. Z. Obermair,  Z. Physik  {\bf 234} 1 (1970).

\bibitem{note1} This assumption is easy to validate for a given Hamiltonian (e.g. the one described in Sec.~\ref{sec:model}). We checked explicitly that the assumption holds for all the models we describe in this paper. 


\bibitem{Magnus64}
W. Magnus,
Comm. Pure and Appl. Math. {\bf 7}, 649 (1964).

\bibitem{note2} 
These simplifications actually carry over to other approaches set up on the level of lowest order diagrammatic approaches, such as self-consistent Hatree-Fock or first order perturbation theory, which, however, are not the subject of study here.

\bibitem{Reed75}
M. Reed and B. Simon, Methods of Modern Mathematical Physics vols 3-4
(Academic Press) (1975).

\bibitem{Barnab05}
X. Barnab\'e-Th\'eriault,  A.  Sedeki,  V.  Meden,  and
K. Sch\"onhammer, Phys. Rev. B {\bf 71}, 205327 (2005).


   
\bibitem{Haldane80}
  F.D.M.~Haldane, Phys.~Rev.~Lett.~{\bf 45}, 1358 (1980).

\bibitem{Kane92}
  C.L.~Kane and M.P.A.~Fisher, Phys.~Rev.~Lett.~{\bf 68}, 1220 (1992).

\bibitem{Meden03}
V.~Meden, S.~Andergassen, W.~Metzner, U.~Schollw\"ock, and K.~Sch\"onhammer, EPL~{\bf 64}, 769 (2003).


\bibitem{Karrasch15}
C. Karrasch and J. E. Moore,
 Phys. Rev. B {\bf 92}, 115108 (2015).


\bibitem{Mitra11}
  A.~Mitra and T.~Giamarchi, Phys.~Rev.~Lett. {\bf 107}, 150602 (2011). 

\bibitem{Mitra12} 
  A.~Mitra and T.~Giamarchi, Phys.~Rev.~B {\bf 85}, 075117 (2012). 

\bibitem{Mitra13} 
  A.~Mitra, Phys.~Rev.~B {\bf 87}, 205109 (2013).

\bibitem{Mitra15} 
M. Schir\'o, A. Mitra,
Phys. Rev. B {\bf 91}, 235126 (2015)

\bibitem{Mitra17}
A.~Mitra, 
arXiv:1703.09740

\end{thebibliography}
\end{document}